\documentclass[aps,prd,10pt,showpacs,amsmath,twocolumn,floatfix,amssymb,nofootinbib,longbibliography]{revtex4-2}
 
\usepackage{graphicx}
\usepackage{comment}
\usepackage[usenames]{color}
\usepackage{bm}
\usepackage{ifpdf}
\usepackage{floatrow}
\usepackage{makecell}
\usepackage{caption}
\usepackage{subcaption}
\usepackage{stackrel}
\usepackage{enumitem}

\usepackage[normalem]{ulem}
\usepackage[dvipsnames]{xcolor}
\usepackage[utf8]{inputenc}
\usepackage{hyperref}
\hypersetup{
  pdfnewwindow=true,      
  colorlinks=true,        
  linkcolor=PineGreen,    
  citecolor=PineGreen,    
  filecolor=PineGreen,    
  urlcolor=PineGreen      
}
\newcommand{\be}{\begin{eqnarray}}
\newcommand{\ee}{\end{eqnarray}}

\usepackage{parskip}
\usepackage{anyfontsize}

\begin{document}

\title{Toward extracting scattering phase shift from integrated correlation functions III: coupled-channels} 

\author{Peng~Guo}
\email{peng.guo@dsu.edu}

\affiliation{College of Arts and Sciences,  Dakota State University, Madison, SD 57042, USA}
\affiliation{Kavli Institute for Theoretical Physics, University of California, Santa Barbara, CA 93106, USA}

\author{Frank~X. Lee}
\email{fxlee@gwu.edu}
\affiliation{Department of Physics, George Washington University, Washington, DC 20052, USA}

\date{\today}

\begin{abstract}
The formalism developed in  Refs.\cite{Guo:2023ecc,Guo:2024zal} that connects   integrated correlation function of a trapped two-particle system to infinite volume scattering phase shift is further extended to coupled-channel systems in the present work. Using a trapped non-relativistic two-channel system as an example, a new relation is derived that retains the same structure as in the single channel, and has explicit dependence on the phase shifts in both channels but not on the inelasticity. The relation is illustrated by a exactly solvable coupled-channel quantum mechanical model with contact interactions. It is further validated by path integral Monte Carlo simulation of a quasi-one-dimensional model that can admit general interaction potentials. In all cases, we found rapid convergence to the infinite volume limit as the trap size is increased, even at short times, making it potentially a good candidate to overcome signal-to-noise issues in Monte Carlo applications.

\end{abstract}

\maketitle

\section{Introduction}\label{sec:intro}

Scattering is a ubiquitous  tool in probing the nature of interactions between particles, either in atomic and molecular physics or in nuclear and particle physics. An effective theoretical approach is the use of a finite box with periodic boundary conditions to enclose the system under consideration. General relations can be established that connect the quantized energies of the system to the infinite-volume scattering phaseshifts, regardless of how the energy spectrum is obtained. The most famous of the approach is the L\"{u}scher formula \cite{Luscher:1990ux} which has been widely applied in the field of lattice QCD to obtain scattering parameters in hadron-hadron interactions~\cite{Rummukainen:1995vs,Christ:2005gi,Bernard:2008ax,He:2005ey,Lage:2009zv,Doring:2011vk,Guo:2012hv,Guo:2013vsa,Kreuzer:2008bi,Polejaeva:2012ut,Hansen:2014eka,Mai:2017bge,Mai:2018djl,Doring:2018xxx,Guo:2016fgl,Guo:2017ism,Guo:2017crd,Guo:2018xbv,Mai:2019fba,Guo:2020wbl,Guo:2018ibd,Guo:2019hih,Guo:2019ogp,Guo:2020kph,Guo:2020spn,Woss2020,Fischer:2020jzp,Hansen:2020otl,blanton2021interactions,mai2021threebody,Bovermann_2019,Lee:2020fbo,Guo:2018zss,Culver:2019qtx,Alexandru:2020xqf,Brett:2021wyd,Briceno:2016mjc,Moir:2016srx,Mai:2021lwb,Rusetsky:2019gyk,Mai:2021nul,Lee:2021kfn}. 
In the case of nucleon-nucleon scattering, in addition to the L\"{u}scher method, there is the potential method by the HALQCD collaboration~\cite{PhysRevLett.99.022001,10.1143/PTP.123.89,PhysRevD.99.014514,ISHII2012437,AokiEPJA2013} also relying on the discrete energy spectrum.  

The nucleon-nucleon system presents unique challenges to such methods.
First, the  signal-to-noise (S/N)  issue \cite{lepage1989analysis,DRISCHLER2021103888}  requires exponentially more  statistics  in stochastic evaluation of the path integral of the two-nucleon correlation. Second, at larger volumes the condensed  energy spectrum requires a large number of interpolating operators to resolve~\cite{Bulava:2019kbi}. The required Euclidean time   to display the signal of clear plateau  could be  well into the region where the signal is swamped by noise. 
The challenges resulted in the so-called ``two-nucleon controversy": The results from the L\"uscher  formula method and HAL QCD potential method do not agree~\cite{DRISCHLER2021103888}. The issue is most likely in  the energy spectrum extraction~\cite{Aoki:2023qih}.   

These challenges motivated alternative approaches in recent years. One is the spectral function based methods that utilize Bayesian inference~\cite{Bulava:2019kbi,Hansen:2019idp,Bailas:2020qmv}. Another is the integrated correlation function method proposed in Refs.~\cite{Guo:2023ecc,Guo:2024zal}. We showed that in a two-particle system, the difference between interacting and non-interacting integrated correlation functions in finite volume can be related to  infinite-volume phase shift through a weighted integral.
It is demonstrated that the relation has a rapid  convergence rate at short Euclidean times, even with a modestly small sized box. The fast convergence at short Euclidean time  makes it  potentially  a good candidate to overcome the S/N problem in two-nucleon systems.  Since the formalism involves correlation functions rather than energy spectrum, it is in principle free from issues encountered at large volumes, such as increasingly dense energy spectrum and the extraction of low-lying states.  

The aim of the present work is to extend the single-channel formalism in  Refs.~\cite{Guo:2023ecc,Guo:2024zal} to beyond the elastic scattering region to include the effect of inelastic coupled-channel dynamics. 
We will limit our discussion to two-particle systems in two coupled channels with non-relativistic dynamics in one spatial dimension and one temporal dimension. Such a system can showcase the main ideas of the formalism while being amenable to testing by  exactly solvable models.
We also carry out Monte Carlo simulations to validate the new relation for coupled channels. 
 
The paper is organized as follows. The single-channel integrated correlation function formalism is outlined briefly in Sec.~\ref{singlechanformula}.  The extension to coupled-channel systems  is presented in  Sec.~\ref{coupledchanformula}.  A quasi-one-dimensional  Monte Carlo test  is presented in  Sec.~\ref{MCquasi1D}.  Summary and outlook are given in Sec.~\ref{summary}. Some technical details are in an appendix.

\section{Summary of single-channel integrated correlation function formalism}\label{singlechanformula}  
 In this section, we outline the basic ingredients needed for an extension of the formalism from single channel to two coupled channels.
 
 In Ref.~\cite{Guo:2023ecc,Guo:2024zal}, we derived a relation that connects the  integrated    correlation functions for  two particles in a trap to the scattering phase shift, $\delta (\epsilon)$, through a weighted integral, 
 \begin{equation}
 C(t) - C_0 (t) \stackrel{\text{trap} \rightarrow \infty}{\rightarrow} \frac{i\,t}{\pi} \int_0^{\infty} d \epsilon  \,\delta(\epsilon)\, e^{ -i\,\epsilon\, t},
 \label{single}
 \end{equation} 
 where $C(t)$ and $C_0 (t)$ are integrated correlation functions for two interacting and non-interacting particles in the trap. 
 The relation is derived in Minkowski time. Its Euclidean time version can be obtained by an analytical continuation ($t\to -i\tau$),
  \begin{equation}
 C(\tau) - C_0 (\tau) \stackrel{\text{trap} \rightarrow \infty}{\rightarrow} \frac{\tau}{\pi} \int_0^{\infty} d \epsilon  \,\delta(\epsilon)\, e^{- \epsilon \tau}.
 \label{single2}
 \end{equation}  
 The integrated two-particle correlation function for non-relativistic systems is defined through summing over all the modes of  two-particle correlation functions  along the diagonal,
\begin{equation}
C(t) = \int d r\, C(r t; r 0),
\end{equation}  
where  $C(r t; r' 0)$ is the expectation value of time-ordered two-particle interpolating operators,  
\begin{equation}
C(r t; r' 0) =    \langle 0 | \mathcal{T} \left [ \widehat{\mathcal{O}} (r, t) \widehat{\mathcal{O}}^\dag (r', 0)  \right ]  | 0 \rangle.
\end{equation}
The  $\widehat{\mathcal{O}}^\dag (r', 0)$ and $\widehat{\mathcal{O}} (r, t)$ denote creation  and annihilation operators  to create two particles with relative coordinate of $r'$ at time $0$ at source, and then annihilate them with relative coordinate of $r$ at time $t$ at sink, respectively.   For instance,  for a relativistic $\phi^4$ lattice model in \cite{Guo:2024zal}, two-particle operator after projecting out center of mass motion in the rest frame is constructed by
\begin{equation}
 \widehat{\mathcal{O}}^\dag (r, t) = \frac{1}{\sqrt{2 L}} \int_0^L d x_2 \phi(r + x_2, t) \phi(x_2 , t),
\end{equation}
where $\phi(x,t)$ is the field operator of a scalar particle, and $L$ stands for the length of a periodic box. See Refs.~\cite{Guo:2023ecc,Guo:2024zal} for more examples and technical details in  both non-relativistic  and relativistic systems. 

On the other hand, two-particle correlation functions can be expressed in terms of wavefunctions in the spectral representation. For example,  inserting complete energy basis in between two-particle creation and annihilation operators: $\sum_\epsilon | \epsilon \rangle \langle \epsilon | =1$, and defining two-particle wave function  by
\begin{equation}
 \langle \epsilon |  \widehat{\mathcal{O}}^\dag (r, 0) | 0 \rangle  = \psi_\epsilon^* (r) ,
\end{equation} the forward time propagating non-relativistic two-particle correlation function is given by~\cite{Guo:2023ecc}, 
\begin{equation}
C(r t; r' 0)|_{t>0} =  \sum_\epsilon \psi_\epsilon (r) \psi^*_\epsilon (r')  e^{- i \epsilon t}  ,
\label{Crr}
\end{equation}
where the two-particle relative wavefunction for trapped systems satisfies the Schr\"odinger equation,
\begin{equation}
\left  [ -\frac{1}{2 \mu} \frac{d^2 }{d r^2}  + U_{trap} (r) +V (r) \right ] \psi_\epsilon (r) = \epsilon \psi_\epsilon (r).
\end{equation}
 We use a unit system with $\hbar=c=1$.
The $\mu$ denotes the reduced mass of the two-particle system,  $U_{trap} (r)$ trap potential, and $V (r)$ two-particle interaction potential. In this representation, the integrated  version of Eq.\eqref{Crr} is simply related to the energy spectrum by,
 \begin{equation}
C(t)   =  \sum_\epsilon  e^{-  i\,\epsilon \, t }  . \label{Cenergyspectrarep}
\end{equation}
The energy spectrum of trapped interacting systems can be determined by L\"uscher formula-like quantization conditions, see e.g. discussions in Ref.~\cite{Guo:2023ecc,Guo:2024zal}.

To connect the integrated two-particle correlation function to scattering phase shift in infinite volume, we use its Green's function representation~\cite{Guo:2023ecc,Guo:2024zal},
\begin{equation}
C(t) = i \int_{-\infty}^\infty \frac{d \lambda}{2\pi} Tr \left [ G^{(trap)} ( \lambda + i 0) \right ] e^{- i \lambda t}, \label{CdefGreen}
\end{equation} 
where the $i 0$ notation means that the integration variable  approaches the real axis from above. The Green's function is the solution of Dyson equation, see e.g. Eq.(26) in Ref.\cite{Guo:2022row}.   Similarly  the two-particle correlation function is connected to Green's function by 
\begin{equation}
C(rt ; r' 0) = i \int_{-\infty}^\infty \frac{d \lambda}{2\pi}  G^{(trap)} (  r,r' ; \lambda + i 0)   e^{- i \lambda t}.  
\end{equation} 
The spectral representation of Green's function  without contributions from anti-particle time-backward propagation has the form, 
\begin{equation}
G^{(trap)} ( \lambda ) = \sum_{\epsilon} \frac{\psi_\epsilon (r) \psi^*_\epsilon (r') }{\epsilon - \lambda } .
\end{equation} 
The trace for non-relativistic trapped systems is defined by,
\begin{equation}
 Tr \left [ G^{(trap)} ( \lambda  ) \right ]  = \int d r G^{(trap)} (r,r ; \lambda  ).
\end{equation} 
Using the relation between trace of Green's functions and scattering phase shift through a dispersion integral  in infinite volume, see e.g. Refs.~\cite{Guo:2022row,Guo:2023ecc,Guo:2024bar},
\begin{align} 
& Tr  \left [ G^{(\infty)} (  \epsilon) -G_0^{(\infty)} (    \epsilon )   \right ]  = - \frac{1}{\pi}  \int_{0}^{\infty} d \lambda  \frac{\delta (\lambda)}{ (\lambda - \epsilon)^2},\label{Ginfphaseshift}
\end{align}
we arrive at the final relation in Eq.\eqref{single} by,
\begin{align}
& C(t) - C_0 (t) = \sum_l \left [  e^{-  i\epsilon_l t } -    e^{- i \epsilon_l^{(0)} t  } \right ]    \nonumber \\
&  \stackrel{\text{trap} \rightarrow \infty}{\rightarrow} i \int_{-\infty}^\infty \frac{d \lambda}{2\pi} Tr \left [ G^{(\infty)} ( \lambda + i 0) - G^{(\infty)}_0 ( \lambda + i 0) \right ] e^{- i \lambda t} \nonumber \\
& = \frac{i\,t}{\pi} \int_0^{\infty} d \epsilon  \,\delta(\epsilon)\, e^{- i\epsilon t} . 
\label{s1}
\end{align} 
The $\epsilon_l $ and $\epsilon_l^{(0)}$ refer to eigen-energy of interacting and non-interacting particles systems in the trap, respectively.

\section{Coupled-channel extension of integrated correlation function formalism}\label{coupledchanformula}  

The integrated correlation function formalism toward extracting  scattering dynamics can  in principle be generalized to include coupled-channel inelastic effects in a relatively straightforward manner.   We start with Green's function and its relation to phase shifts that has been worked out in Ref.~\cite{Guo:2024bar} for coupled channels.  To illustrate the main ideas,
we will only consider non-relativistic dynamics and two coupled channels in this work.  Such a system in an external trap may be described by the Hamiltonian matrix,
\begin{equation}
\hat{H}^{(trap)} = 
\begin{bmatrix}  
 H_1^{(trap)} +V_1 (r) & \gamma (r) \\
 \gamma (r) &  H_2^{(trap)} +V_2 (r) 
  \end{bmatrix},
\end{equation}
where $V_{1,2} (r)$ is the short-range interaction potential in each channel, and $\gamma (r)$ the interaction potential between the two channels. The  $H_{1,2}^{(trap)} $ is non-interacting trapped Hamiltonian in each channel defined by,
\begin{equation}
H_n^{(trap)} = \sigma_n  - \frac{1}{2 \mu_n } \frac{d^2}{ d r^2}  + U_n^{(trap)} (r),\quad n=1,2,
\end{equation}
where $\sigma_n$ and $\mu_n$ are threshold energy  and reduced mass, and $U_n^{(trap)} (r)$ the trapping potential, in each channel.

For a coupled-channel system, the integrated two-particle correlation function can still be defined through the trace of Green's function as in Eq.(\ref{CdefGreen}), except on the right-hand side of Eq.(\ref{CdefGreen}) we have now a matrix of Green's functions.
\begin{equation}
   \mathcal{G}^{(trap)} (r,r' ; \lambda  ) =  
    \begin{bmatrix}  G_{11}^{(trap)} (r,r' ; \lambda  ) & G_{12}^{(trap)} (r,r' ; \lambda  )  \\
    G_{21}^{(trap)} (r,r' ; \lambda  ) & G_{22}^{(trap)} (r,r' ; \lambda  ) 
    \end{bmatrix}  ,
\end{equation} 
It satisfies the matrix version of Dyson equation,
\begin{align}
  &  \mathcal{G}^{(trap)} (r,r' ; \lambda  ) =  \mathcal{G}_0^{(trap)} (r,r' ; \lambda  )   \nonumber \\
   & + \int d r''  \mathcal{G}_0^{(trap)} (r,r'' ; \lambda  )  V(r'')   \mathcal{G}^{(trap)} (r'', r' ; \lambda  ). \label{Dysoneq}
 \end{align}
The non-interacting Green's function matrix is defined by, 
\begin{equation}
   \mathcal{G}_0^{(trap)} (r,r' ; \lambda  ) =  
    \begin{bmatrix}  G_{1}^{( trap, 0)} (r,r' ; \lambda  ) &  0  \\
     0 & G_{2}^{( trap, 0)} (r,r' ; \lambda  ) 
    \end{bmatrix}  ,
\end{equation} 
and the potential matrix by,
\begin{equation}
   V (r   ) =  
  \begin{bmatrix}  
 V_1 (r) & \gamma (r) \\
 \gamma (r) &   V_2 (r) 
  \end{bmatrix}  .
\end{equation} 
The individual channel non-interacting Green's functions in the trap are the solutions of differential equation,
\begin{align}
& \left  [ \lambda - \sigma_n  + \frac{1}{2 \mu_n} \frac{d^2 }{d r^2}  - U_n^{(trap)} (r)   \right ] G_n^{(trap, 0)} (r,r' ; \lambda  )  \nonumber \\
&= \delta(r-r') .
\end{align}

The definition of trace of Green's function matrix must include the trace of the matrix as well. Hence, only diagonal terms of Green's function matrix contribute in  the integrated two-particle correlation function,
\begin{equation}
C(t) = i \int_{-\infty}^\infty \frac{d \lambda}{2\pi}   \left [  \sum_{n =1}^2  \int d r\, G_{nn}^{(trap)} (r,r;  \lambda + i 0) \right ] e^{- i \lambda t} . \label{integratedCdef}
\end{equation} 

In the following two subsections, we will connect the difference of integrated correlation functions between interacting and non-interacting particles systems and phase shifts in the coupled channels as the infinite volume limit is approached, just as in single-channel cases.

\subsection{Coupled-channel scattering solutions in infinite volume}
As detailed in Ref.~\cite{Guo:2024bar}, in the case of short-range interaction, the infinite volume scattering solutions for coupled-channel dynamics can be worked out straightforwardly. Here, we briefly outline  the main results. The scattering coupled-channel wavefunctions can be determined by Lippmann-Schwinger equations,
 \begin{align}
 &   \begin{bmatrix}    \psi^{(\infty)}_1 (r; E)  \\  \psi^{(\infty)}_2 (r; E)   \end{bmatrix}   
  =  \begin{bmatrix}    \psi^{(\infty, 0)}_1 (r; E)  \\  \psi^{(\infty , 0)}_2 (r;E)   \end{bmatrix}   \nonumber \\
  &+  \int d r'  \begin{bmatrix}    G_1^{(\infty, 0)}   (r-r'; E)  & 0   \\   0 & G_2^{(\infty , 0)}   (r-r'; E)   \end{bmatrix}    \nonumber \\
  & \times    \begin{bmatrix}  
 V_1 (r') & \gamma (r') \\
 \gamma (r') &   V_2 (r') 
  \end{bmatrix}        \begin{bmatrix}    \psi^{(\infty)}_1 (r';E)  \\  \psi^{(\infty)}_2 (r'; E)   \end{bmatrix}  ,
 \end{align}
 where the infinite volume free-particle Green's function in individual channel is defined by
 \begin{align}
 & G_n^{(\infty, 0)}   (x; E + i 0)  \nonumber \\
 &=  \int \frac{d p}{2\pi} \frac{e^{i p x}}{ E -   \sigma_n - \frac{p^2}{2 \mu_n } + i 0}  = - \frac{i \mu_n}{ k_n} e^{i k_n |x|} .
 \end{align}
 The total energy is related to the threshold energy and and relative momentum  in each channel by the dispersion relation,
  \begin{equation}
 E =  \sigma_n + \frac{k_n^2}{2\mu_n}, \ \ n=1,2 .
 \label{totalE}
 \end{equation}

The scattering amplitudes are defined through the asymptotic behavior of wavefunction solutions by considering two sets of independent  boundary conditions of incoming waves:
 \begin{align}
(1)  \begin{bmatrix}    \psi^{(\infty, 0)}_1 (r;E)  \\  \psi^{(\infty, 0)}_2 (r;E)   \end{bmatrix}  &=  \begin{bmatrix}   e^{ i  k_1 r}  \\  0  \end{bmatrix} , \nonumber \\
(2)   \begin{bmatrix}    \psi^{(\infty, 0)}_1 (r;E)  \\  \psi^{(\infty , 0)}_2 (r;E)   \end{bmatrix} & =    \begin{bmatrix}   0  \\   e^{ i k_2 r }  \end{bmatrix} .
 \end{align}
Under these conditions, the asymptotic solutions of wavefunctions can be written as,
  \begin{align}
(1)  &  \begin{bmatrix}    \psi^{(\infty)}_1 (r;E)  \\  \psi^{(\infty)}_2 (r;E)   \end{bmatrix}  \stackrel{|r| \rightarrow \infty}{\rightarrow}  \begin{bmatrix}   e^{ i  k_1 r}  + i \frac{\mu_1}{k_1 } T_{11} e^{i k_1 |r|}   \\   i \frac{\mu_2}{k_2 } T_{21} e^{i k_2 |r|}    \end{bmatrix} ,  \nonumber \\
 (2)  &  \begin{bmatrix}    \psi^{(\infty)}_1 (r;E )  \\  \psi^{(\infty)}_2 (r;E)   \end{bmatrix}  \stackrel{|r| \rightarrow \infty}{\rightarrow}    \begin{bmatrix}   i \frac{\mu_1}{k_1 } T_{12} e^{i k_1 |r|}  \\  e^{ i  k_2 r}  +  i \frac{\mu_2}{k_2 } T_{22} e^{i k_2 |r|}    \end{bmatrix} .
 \end{align}
 The coupled-channel scattering amplitudes (T-matrix elements) can be parameterized by two phase shifts, $\delta_{1,2}$, and one inelasticity, $\eta \in [0,1]$, see e.g. \cite{Guo:2010gx,Guo:2011aa,Guo:2012hv,Guo:2013vsa,Guo:2021uig},
 \begin{align}
T_{11}  &=  \frac{k_1}{\mu_1} \frac{\eta e^{2 i \delta_1} -1}{2i} , \ \ \ \  T_{22}  =  \frac{k_2}{\mu_2} \frac{\eta e^{2 i \delta_2} -1}{2i} ,  \nonumber \\
T_{12}  & =  T_{21} = \sqrt{  \frac{k_1}{\mu_1}  \frac{k_2}{\mu_2} }  \frac{ \sqrt{1- \eta^2 } e^{i  ( \delta_1 + \delta_2)} }{2} . \label{Tmatparam}
 \end{align}
 We remark that short-range interactions have been assumed, so only even-parity solutions are contributing. The $S$-matrix  is defined by, see e.g. Ref.~\cite{Guo:2024bar},
  \begin{align}
 S(E)   & =   \begin{bmatrix} 1+ 2 i  \frac{\mu_1}{k_1}   T_{11}  &2  i  \sqrt{  \frac{\mu_1}{k_1}  \frac{\mu_2}{k_2}  } T_{12}   \\  2 i   \sqrt{  \frac{\mu_1}{k_1}  \frac{\mu_2}{k_2}  }  T_{21} & 1+ 2 i  \frac{\mu_2}{k_2}   T_{22}  \end{bmatrix}   \nonumber \\
&  =      \begin{bmatrix}   \eta e^{2 i \delta_1}    & i       \sqrt{1- \eta^2 } e^{i  ( \delta_1 + \delta_2)}    \\   i      \sqrt{1- \eta^2 } e^{i  ( \delta_1 + \delta_2)}   &    \eta e^{2 i \delta_2}    \end{bmatrix}  ,
 \end{align}
 and   it satisfies the unitarity relation  $S^\dag (E) S(E) =1$.   

The infinite volume Green's functions for coupled-channel system satisfy the similar matrix version of Dyson  equation as in trapped systems.  As shown in Ref.~\cite{Guo:2024bar}, the trace of the difference of integrated Green's function between interacting and non-interacting systems is related to two scattering phase shifts by a dispersion relation, also known as Krein's theorem  \cite{zbMATH03313022}  in spectral theory,
\begin{align} 
& Tr  \left [ \mathcal{G}^{(\infty)} (  E) -\mathcal{G}_0^{(\infty)} (    E)   \right ] \nonumber \\
& =  \sum_{n=1}^2 \int_{-\infty}^{\infty} d r \left [ G_{nn}^{(\infty)} (r,r ;  E) -G_{n}^{(\infty, 0)} ( 0 ;  E)   \right ] \nonumber \\
&  = - \frac{1}{\pi}  \int_{\sigma_1}^{\infty} d \lambda  \frac{\delta_1 (\lambda)}{ (\lambda - E)^2} - \frac{1}{\pi}  \int_{\sigma_2}^{\infty} d \lambda  \frac{\delta_2 (\lambda)}{ ( \lambda - E )^2}, \label{Ginfphaseshift2}
\end{align}
whose absorptive part, 
\begin{align}
& Im \left [ Tr  \left [ \mathcal{G}^{(\infty)} (  E) -\mathcal{G}_0^{(\infty)} (    E)   \right ] \right ] \nonumber \\
 & =  \sum_{n=1}^2 \int_{-\infty}^{\infty} d r Im \left [ G_{nn}^{(\infty)} (r,r ;  E) -G_{n}^{(\infty, 0)} ( 0 ;  E)   \right ]   \nonumber \\
    & = -   \frac{d  }{d E}  \left [   \delta_1 (E)  + \delta_2 (E) \right ]   ,\label{ImGinfphaseshift}
\end{align}
gives the coupled-channel version of Friedel formula \cite{doi:10.1080/00018735400101233,Friedel1958,Faulkner_1977} in condensed matter theory, also see discussions in e.g. Ref.~\cite{Guo:2022row}. 
Combining the definition of $C(t)$ in Eq.\eqref{integratedCdef} and Eq.\eqref{Ginfphaseshift2} yields the final relation in Minkowski space,
 \begin{equation}
 C(t) - C_0 (t) \stackrel{\text{trap} \rightarrow \infty}{\rightarrow} \frac{i\, t}{\pi} \left [ \int_{\sigma_1}^{\infty}   \delta_1(\epsilon)   + \int_{\sigma_2}^{\infty}     \delta_2(\epsilon)  \right ]  e^{-i\, \epsilon\, t}  d \epsilon , \label{mainresult1}
 \end{equation}  
 The corresponding relation in Euclidean space is given by,
  \begin{equation}
 C(\tau) - C_0 (\tau) \stackrel{\text{trap} \rightarrow \infty}{\rightarrow} \frac{\tau}{\pi} \left [ \int_{\sigma_1}^{\infty}   \delta_1(\epsilon)   + \int_{\sigma_2}^{\infty}     \delta_2(\epsilon)  \right ]  e^{- \epsilon \tau}  d \epsilon . \label{mainresult}
 \end{equation}  
 Eq.\eqref{mainresult} is the main result of this work. In the following, we will examine its properties from multiple angles.

\subsection{Energy spectrum representation of integrated correlation function and quantization condition} 
For trapped systems, similar to the single-channel spectral representation, see e.g. Ref.~\cite{Guo:2023ecc},  the spectral representation of full Green's function matrix   is given by,
\begin{equation}
\mathcal{G}^{(trap)} (r,r' ; \lambda) = \sum_{l} \frac{\psi^{(trap)} (r;  \epsilon_l)  \psi^{(trap)\dag} (r' ; \epsilon_l ) }{\lambda - \epsilon_l}, 
\end{equation} 
where  
\begin{equation}
\psi^{(trap)} (r;  \epsilon  ) =   \begin{bmatrix} \psi^{(trap)}_1 (r;  \epsilon ) \\ \psi^{(trap)}_2 (r;  \epsilon  ) \end{bmatrix} , 
\end{equation}
stands for a column matrix of coupled-channel  wavefunctions,  and is  the eigen-solution of coupled-channel Schr\"odinger equation,
\begin{equation}
\hat{H}^{(trap)} \psi^{(trap)} (r;  \epsilon_l)  = \epsilon_l  \psi^{(trap)} (r;  \epsilon_l) .  \label{Schrodingertrap}
\end{equation}
The normalization condition for the wavefunctions of coupled-channel systems is given by
\begin{equation}
\int d r \psi^{(trap)\dag} (r ; \epsilon_l ) \psi^{(trap) } (r ; \epsilon_{l'}) = \delta_{l,l'},
\end{equation}
which leads to the trace,
\begin{equation} 
Tr \left [\mathcal{G}^{(trap)} (  \lambda)  \right ] = \sum_{n=1}^2  \int d r  G_{nn}^{(trap)} (r,r ;  \lambda)    = \sum_{l} \frac{1 }{\lambda - \epsilon_l}.
\end{equation} 
Using the definition of  the integrated correlation function for coupled-channel system in Eq.(\ref{integratedCdef}), we thus find the energy spectrum representation of integrated correlation function for coupled-channel system:  
\begin{equation} 
C(t)  =  \sum_{l} e^{- i\,\epsilon_l \, t}.
\label{Cenergyspectrarep-coupled}
\end{equation}
It resembles the result in single-channel systems in Eq.(\ref{Cenergyspectrarep}), but now $\epsilon$ stands for the discrete eigen-energies in coupled-channel systems. Therefore, similar to relation in Eq.(\ref{s1}) in single-channel cases,  we find a useful relation,
 \begin{align}
&  \sum_l \left [  e^{- i\, \epsilon_l \,t } -    e^{- i\, \epsilon_l^{(0)} \, t } \right ]  \nonumber \\
& \stackrel{\text{trap} \rightarrow \infty}{\rightarrow} \frac{i \, t}{\pi}  \left [ \int_{\sigma_1}^{\infty}   \delta_1(\epsilon)   + \int_{\sigma_2}^{\infty}     \delta_2(\epsilon)  \right ]  e^{- i\,\epsilon  \, t }  d \epsilon, \label{sumenergyphaseshiftcouplechannel}
 \end{align}  
 for coupled-channel systems in terms of energy spectra.

The  eigen-energy of coupled-channel systems in a trap can be solved by diagonalizing the Hamiltonian in the basis of non-interacting trapped systems. Expanding coupled-channel wavefunctions,
\begin{equation}
\psi^{(trap)} (r;  \epsilon  ) =  \sum_{l}  \begin{bmatrix}  c_l^{(1)} \phi^{(1)}_l (r ) \\ c_l^{(2)}  \phi^{(2)}_l (r   ) \end{bmatrix} ,
\end{equation}
in terms of the eigen-solution of  Schr\"odinger equation for non-interacting trapped systems,
 \begin{equation}
 H_n^{(trap)} \phi^{(n)}_l (r ) = \epsilon_l^{(n,0)} \phi^{(n)}_l (r ), \ \ n=1,2,
\end{equation}
the Schr\"odinger equation in   Eq.(\ref{Schrodingertrap}) is converted to a matrix equation,
\begin{equation}
\sum_{l'}  H^{(trap)}_{l,l'}  \begin{bmatrix}  c_{l'}^{(1)} \\ c_{l'}^{(2)}  \end{bmatrix}_m  = \epsilon_m  \begin{bmatrix}  c_{l}^{(1)} \\ c_{l}^{(2)}  \end{bmatrix}_m,
\end{equation} 
where $\epsilon_m$ is the $m$-th eigen-energy of this equation, different from  $\epsilon_l^{(n,0)}$.
The matrix element of Hamiltonian  in  the basis of non-interacting trapped systems is thus given by (omitting the label trap),
 \begin{equation}
H_{l,l'} = \begin{bmatrix}  
\delta_{l,l'} \epsilon_l^{(1,0)} + V^{(1)}_{l,l'} &  \gamma_{l,l'}  \\
 \gamma^*_{l',l}  & \delta_{l,l'} \epsilon_l^{(2,0)} + V^{(2)}_{l,l'}
\end{bmatrix} ,
\end{equation}
where the matrix elements $V^{(1,2)}_{l, l'}$ and $\gamma_{l, l'}$  are defined by
\begin{align}
V^{(n)}_{l,l'}  &= \int d r   \phi^{(n)*}_l (r ) V_n (r)  \phi^{(n)}_{l'} (r )  ,  \ \ n=1,2,  \nonumber \\
\gamma_{l,l'} &= \int d r \phi^{(2) *}_l (r )  \gamma (r) \phi^{(1)}_{l'} (r )  . \label{Vnnp}
\end{align}

On the other hand, as discussed in Ref.~\cite{Guo:2023ecc,Guo:2024zal}, also see relevant discussions in e.g. Ref.~\cite{Guo:2021lhz,Guo:2021qfu,Guo:2021uig,Guo:2020spn,Guo:2020ikh,Guo:2020kph,Guo:2019hih,Guo:2019ogp},  when the size of the trap is much larger than the range of interaction, the short-range interaction potential can be well approximated by contact interactions,
\begin{equation}
V(r) =  \begin{bmatrix} V_1 \delta (r) & \gamma  \delta (r)  \\ \gamma  \delta (r)  & V_2   \delta (r)  \end{bmatrix} . \label{contactpot}
\end{equation}
The quantization condition that determines  the discrete eigen-energy  for trapped systems  can be derived through homogeneous Lippmann-Schwinger equations~\cite{Guo:2021lhz,Guo:2021qfu,Guo:2021uig}, which  reads in our notation,
\begin{align}
& \det \Bigg [ \begin{bmatrix} V_1 & \gamma \\ \gamma & V_2   \end{bmatrix}^{-1}  - \begin{bmatrix}  G_1^{(trap,0)} (0,0; \epsilon) & 0 \\ 0  & G_2^{(trap,0)} (0,0; \epsilon)    \end{bmatrix}   \Bigg ] \nonumber \\
&  =0. \label{QCVG}
\end{align}
The relation between interaction potentials and infinite volume scattering amplitudes matrix can be readily established with contact interaction approximation~\cite{Guo:2021lhz,Guo:2021qfu,Guo:2021uig},
\begin{equation}
 -  \begin{bmatrix} V_1 & \gamma \\ \gamma & V_2   \end{bmatrix}^{-1}  = T^{-1}  +  \begin{bmatrix}  \frac{i \mu_1}{k_1} & 0 \\  0 & \frac{i \mu_2}{k_2}   \end{bmatrix}    . \label{VTG0infrelation}
\end{equation}
Eliminating interaction potential matrix in Eq.(\ref{QCVG}) and Eq.(\ref{VTG0infrelation}), and also using phase shift and inelasticity parametrization of $T$-matrix in Eq.(\ref{Tmatparam}), the quantization condition for coupled-channel system can be written in a compact form,
\begin{align}
& \eta (1 + g_1 g_2 ) \cos (\delta_1 - \delta_2) + (1- g_1 g_2) \cos (\delta_1 + \delta_2) \nonumber \\
&- \eta ( g_1  - g_2 ) \sin (\delta_1 - \delta_2) - ( g_1 + g_2) \sin (\delta_1 + \delta_2) =0, \label{QCcompact}
\end{align}
where the $g_{1,2}$ functions are defined by
\begin{equation}
g_n =  - \frac{k_n}{\mu_n} G^{(trap, 0)}_0 (0, 0; \epsilon), \ \ n=1,2 .
\end{equation}
The quantization condition in Eq.(\ref{QCcompact}) has the same structure as in Eq.(25) in Ref.~\cite{Guo:2012hv} for finite volume coupled-channel systems and Eq.(51) in  Ref.~\cite{Guo:2021uig} for coupled-channel systems in a harmonic oscillator trap, providing an independent confirmation of our formalism. The $g_{1,2}$ functions play the role of  L\"uscher  zeta function in the finite volume system \cite{Luscher:1990ux}. Explicitly in 1D periodic box of size $L$, the  functions are~\cite{Guo:2016fgl,Guo:2017ism,Guo:2017crd,Guo:2018xbv},
\begin{equation}
g_n = - \cot \left( \frac{L}{2}  \sqrt{2\mu_n (\epsilon-\sigma_n)}\right).
\end{equation}
For the harmonic trap with angular frequency $\omega$, explicit expressions for $g_{1,2}$ functions are given by
\begin{equation}
g_n =  \sqrt{ \frac{ \epsilon - \sigma_n}{2 \omega} }  \frac{\Gamma ( \frac{1}{4} -  \frac{ \epsilon - \sigma_n}{2 \omega})}{\Gamma ( \frac{3}{4} -  \frac{ \epsilon - \sigma_n}{2 \omega})} ,
\end{equation}
which is the periodic function known as  BERW formula in single channel~\cite{Busch98,Guo:2021uig,Guo:2021lhz}.  When the coupling between the two channels is turned off ($\gamma=0$),  the coupled-channel quantization condition in Eq.(\ref{QCcompact}) is reduced to the well-known single-channel L\"uscher formula:  $\cot \delta_n  = g_n$, see e.g. Refs.~\cite{Luscher:1990ux,Busch98,Guo:2021uig,Guo:2021lhz}.

\begin{figure}
 \centering
\includegraphics[width=0.95\textwidth]{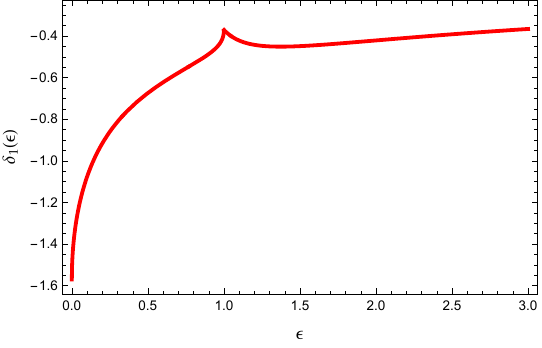}
\includegraphics[width=0.95\textwidth]{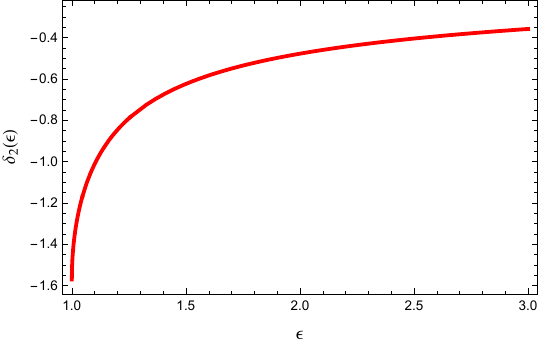}
\includegraphics[width=0.95\textwidth]{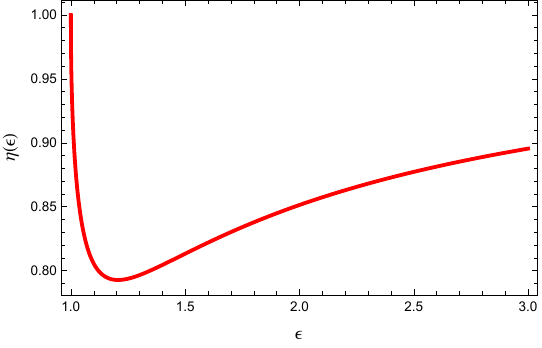}
\captionsetup{singlelinecheck=false, justification=raggedright}
 \caption{Model parameters for phase shifts ($\delta_1$ and $\delta_2 $) and inelasticity ($ \eta$) as defined in Eq.\eqref{phaseinelastexpression}.   The kink in $\delta_1$ is a threshold effect corresponding to the dip in $ \eta$. Other  parameters are taken as:       $V_1=1.0$, $V_2 =0.8$, $\gamma=0.6$, $\sigma_1 =0$, $\sigma_2 =1$ and $\mu_1 = \mu_2 =1$.  }
 \label{phaseinelasticityplots} 
 \end{figure}

\subsection{Example of an exactly solvable model}\label{exactmodel}
Having obtained the relation in Eq.\eqref{mainresult}, we want to check its convergence 
by an exactly solvable model.
With the contact interaction potentials in Eq.(\ref{contactpot}), the analytic solutions of infinite volume scattering amplitudes are available via Eq.(\ref{VTG0infrelation}):
\begin{equation}
T =  -  \begin{bmatrix} V_1 & \gamma \\ \gamma & V_2   \end{bmatrix}      \begin{bmatrix} 1 + \frac{i \mu_1}{k_1}  V_1 &   \frac{i \mu_1}{k_1} \gamma \\   \frac{i \mu_2}{k_2}  \gamma &  1+  \frac{i \mu_2}{k_2} V_2   \end{bmatrix}^{-1}   . 
\end{equation}
The T-matrix elements are given by,
\begin{align}
T_{11}  (\epsilon) &=   \frac{ k_1}{\mu_1}   \frac{    i  ( \gamma^2- V_1 V_2)  -      V_1   \frac{k_2}{\mu_2}      }{(\frac{k_1}{\mu_1}  + i V_1) (\frac{k_2}{\mu_2}  + i V_2) + \gamma^2 }  ,  \nonumber \\
T_{12} (\epsilon)  & = T_{21} (\epsilon) =     \frac{     - \gamma  \frac{ k_1}{\mu_1}       \frac{k_2}{\mu_2}      }{(\frac{k_1}{\mu_1}  + i V_1) (\frac{k_2}{\mu_2}  + i V_2) + \gamma^2 }  ,  \nonumber \\
T_{22}(\epsilon)   &=   \frac{ k_2}{\mu_2}   \frac{    i  ( \gamma^2- V_1 V_2)  -      V_2   \frac{k_1}{\mu_1}      }{(\frac{k_1}{\mu_1}  + i V_1) (\frac{k_2}{\mu_2}  + i V_2) + \gamma^2 }     , \label{Tscatamps}
\end{align}
where $k_n = \sqrt{ 2 \mu_n (\epsilon - \sigma_n) } $. The $S$-matrix elements are,
\begin{align}
S_{11}  (\epsilon) &=      \frac{  (\frac{k_1}{\mu_1}  - i V_1) (\frac{k_2}{\mu_2}  + i V_2) - \gamma^2   }{(\frac{k_1}{\mu_1}  + i V_1) (\frac{k_2}{\mu_2}  + i V_2) + \gamma^2 }  ,  \nonumber \\
S_{12} (\epsilon)  & = S_{21} (\epsilon) =     \frac{     - 2 i\gamma  \sqrt{ \frac{ k_1}{\mu_1}       \frac{k_2}{\mu_2}  }    }{(\frac{k_1}{\mu_1}  + i V_1) (\frac{k_2}{\mu_2}  + i V_2) + \gamma^2 }  ,  \nonumber \\
S_{22}(\epsilon)   &=      \frac{  (\frac{k_1}{\mu_1}  + i V_1) (\frac{k_2}{\mu_2}  - i V_2) - \gamma^2   }{(\frac{k_1}{\mu_1}  + i V_1) (\frac{k_2}{\mu_2}  + i V_2) + \gamma^2 }     . \label{Smatrix}
\end{align}
The scattering amplitudes satisfy the unitarity relation,
\begin{equation}
Im \left [ T_{n n'} (\epsilon)  \right ] = \sum_{l=1}^2 \theta (\epsilon -\sigma_l ) \frac{\mu_l }{k_l} T^*_{n l } (\epsilon)  T_{l n'} (\epsilon)  .
\end{equation}
 The inelasticity and two-channel phase shifts can thus be computed by
 \begin{align}
 \eta (\epsilon) &=
  \begin{cases}
  \sqrt{1-  4 \frac{\mu_1}{k_1}  \frac{\mu_2}{k_2} | T_{12} (\epsilon)  |^2 } , & if \ \ \epsilon > \sigma_2, \\ 
  1, & \text{otherwise},
  \end{cases}
   \nonumber \\
  \delta_n  (\epsilon)& =  
   \begin{cases} 
   \frac{1}{2 i} \ln \left [ \frac{1+ 2 i \frac{\mu_n}{k_n} T_{nn} (\epsilon)  }{\eta (\epsilon)  } \right] , & if \ \ \epsilon > \sigma_n, \\ 
  0, & \text{otherwise}.
  \end{cases} \label{phaseinelastexpression}
 \end{align}
 For the contact interaction in 1D, the phase shifts approach $ - \frac{\pi}{2}$ at thresholds:  $  \delta_n  (\sigma_n) =- \frac{\pi}{2} $. An example for the real parts of the phase shifts and inelasticity as a function of energy is plotted in Fig.~\ref{phaseinelasticityplots}.

 \begin{figure}
\centering\includegraphics[width=0.98\textwidth]{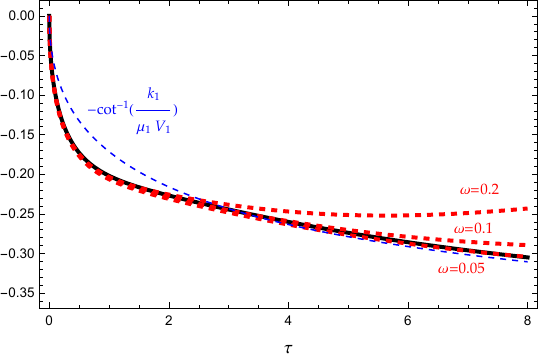}
\caption{Convergence check of Eq.\eqref{mainresult} in an exactly solvable model. The dashed red lines are the difference $C (t) -C_0 (t) $ on the left-hand-side in harmonic oscillator trap with $\omega=0.05, 0.1,0.2$. The solid black curve is  the full result in infinite volume limit on the right-hand-side.  The parameters are taken as:     $V_1=0.6$, $V_2 =0.3$,   $\gamma=0.2$, $\sigma_1 =0$, $\sigma_2 =1$,  and $\mu_1 = \mu_2 =1$.  
The dashed blue curve is the   $C(t)-C_0(t) $ of channel-1 when the inelastic effect is turned off $(V_2=0, \gamma=0)$. }\label{dztHOplot}
\end{figure}

If harmonic oscillator trap is used , the basis functions are explicitly given by,
 \begin{equation}
   \phi^{(n)}_l(r)  = \frac{1}{\sqrt{2^l l!}} \left ( \frac{\mu_{n} \omega}{\pi} \right )^{\frac{1}{4}} e^{- \frac{\mu_{n} \omega}{2} r^2} H_l (\sqrt{\mu_{n} \omega} r), \ n=1,2.
   \label{eq:ho-basis}
\end{equation}
with associated energies $ \epsilon_l^{(\omega)} =   \omega (l+ \frac{1}{2})$.
Hence,  the Hamiltonian matrix for coupled-channel   system with contact interactions in harmonic oscillator trap is given by,
 \begin{align}
 & H_{l,l'} =\nonumber \\
 &  \begin{bmatrix}  
\delta_{l,l'}   ( \sigma_1 +  \epsilon_l^{(\omega)} )   + V^{(1)}_{l,l'} &  \gamma_{l,l'}  \\
 \gamma_{l',l}  & \delta_{l,l'}  ( \sigma_2 + \epsilon_l^{(\omega)})  + V^{(2)}_{l,l'}
\end{bmatrix} , \label{HmatHOtrap}
\end{align}
where matrix elements of potential terms can be evaluated by Eq.\eqref{Vnnp}.
The interacting trapped integrated correlation function in Euclidean time, see Eq.\eqref{Cenergyspectrarep-coupled},
\begin{equation}
    C(\tau ) = \sum_{l}  e^{- E_l \tau},
    \label{Ctau2}
\end{equation} 
can be computed by using energy spectrum $\{E_l\}$   by diagonalizing the Hamiltonian in Eq.\eqref{HmatHOtrap}.
The non-interacting version is given by,
\begin{equation}
C_0 (\tau) = (e^{-\sigma_1 \tau} + e^{-\sigma_2 \tau} ) \frac{1}{2} \mbox{csch} (\frac{\omega \tau}{2}),
\label{Czero2}
\end{equation}
where we have used the identity: $$\sum_l e^{- \omega (l+ \frac{1}{2}) \tau} = \frac{1}{2} \mbox{csch} (\frac{\omega \tau}{2}).$$
Eq.\eqref{Ctau2} and Eq.\eqref{Czero2} will be further elucidated in the path integral representation below.

A numerical demonstration  of the relation in Eq.(\ref{mainresult}) is shown in Fig.~\ref{dztHOplot} using the coupled-channel parametrization in Fig.~\ref{phaseinelasticityplots}. The $C(\tau)$ is computed by feeding $\{E_l\}$ into Eq.\eqref{Ctau2} as mentioned above. The $C_0(\tau)$ is computed by Eq.\eqref{Czero2}. The results with various size of harmonic oscillator traps, $\omega  \in [ 0.005,0.1,0.2]$, are compared with its infinite volume limit on the right-hand-side using the phase shifts in Eq.\eqref{phaseinelastexpression}. We see that as $\omega\to 0$ (which corresponds to $\text{trap}\to \infty$), the relation is satisfied. Even with relative large trap size of $\omega \sim 0.2$,   the difference of integrated correlation functions shows a rapid convergence  to its infinite volume limit  near short times.  To highlight the inelastic effect, the    $C(t)-C_0(t) $  of channel-1 from turning off inelastic dynamics $(V_2 = \gamma =0)$, 
\begin{equation}
    \delta_1 \stackrel{ (V_2  , \gamma ) \rightarrow 0}{ \rightarrow } - \cot^{-1} (\frac{k_1}{\mu_1 V_1}),
\end{equation} 
is also plotted in Fig.~\ref{dztHOplot}. Sizable inelastic effect is observed at short times.

\section{Monte Carlo simulation of a quasi-one-dimensional coupled-channel model}\label{MCquasi1D}

In this section, we propose a quasi-one-dimensional quantum mechanical model to further validate the coupled-channel dynamics when the system is not exactly solvable. Here `quasi' means that the coupled-channel 1D dynamics is embedded in a two-dimensional (2D) waveguide model~\cite{Guo:2024bar}.  In the following, we will show that when the particles are confined in a 2D trap, the number of bound states in the transverse direction can be manipulated in the path integral representation and its degrees of freedom can be integrated out, reducing the 2D system to a quasi-1D coupled-channel quantum mechanical system, which can be used to conduct nontrivial tests on the proposed coupled-channel formalism.

 \begin{figure}[h!]
\centering\includegraphics[width=0.97\textwidth]{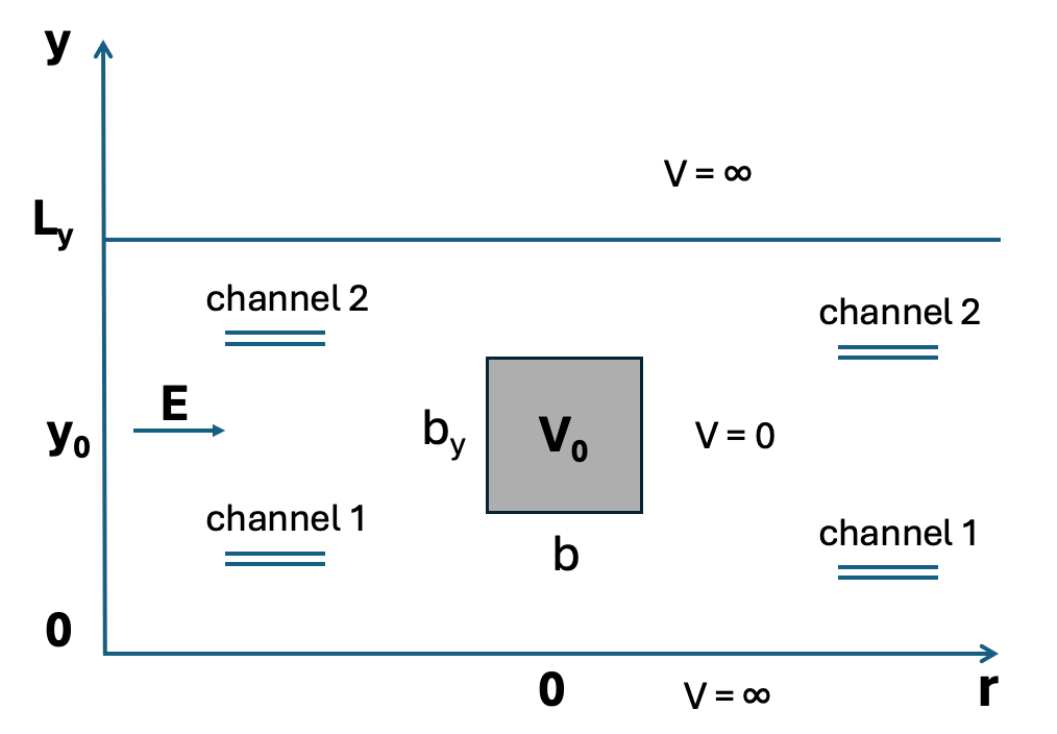}
\caption{Illustration  of quasi-one-dimensional coupled-channel system that is embedded in a 2D geometry. The grey square in the middle represents the 2D square-well interaction potential placed at $(0,y_0)$ that yields the coupled-channel interaction potentials in quasi-1D. Each channel is a two-particle system with total energy $E$ and individual ones as related in Eq.\eqref{totalE}. The system is confined in y-direction by a hard-wall trap of width $L_y$ and a harmonic oscillator trap in $r$-direction. }\label{quasi1Dplot}
\end{figure}

\subsection{Quasi-1D quantum mechanical model} 
Following Ref.~\cite{Guo:2024bar}, the quasi-1D system can be described by a 2D   Hamiltonian,
\begin{equation}
\hat{H} = \hat{H}_r^{(\omega)}  + \hat{H}_y^{(trap)}  + V(r,y),
\end{equation}
where  
\begin{equation}
\hat{H}_r^{(\omega)}  =  - \frac{1}{2 m} \frac{\partial^2}{\partial r^2} +  \frac{1}{2} m \omega^2 r^2 ,
\end{equation}
describes a harmonic oscillator trap in $r$-direction, and,
\begin{equation}
\hat{H}_y^{(trap)}  =  - \frac{1}{2 m} \frac{\partial^2}{\partial y^2} + U^{(trap)}_y (y), 
\end{equation}
a trap in $y$-direction whose form can be chosen. The $V(r,y)$ is used to simulate the short-range interaction between the two particles. 
In this work, we use a simple 2D square-well potential,
\begin{equation}
V(r,y) = 
\begin{cases}
\frac{V_0}{b b_y}  , & if \  r \in [ - \frac{b}{2}, \frac{b}{2}]  \& y \in [ y_0 - \frac{b_y}{2}, y_0 + \frac{b_y}{2}] ,  \\
0, & \text{otherwise},
\end{cases}
\end{equation}
where $b$ and $b_y$ are the width of square well along $r$- and $y$-direction respectively. The center of the square well is placed at $(0, y_0)$. The schematic diagram with a hard-wall trap in $y$-direction is shown in Fig.~\ref{quasi1Dplot}.  At the limit of shrinking width of square well from both directions, the above potential is reduced to a 2D $\delta$-potential,
\begin{equation}
V(r,y)  \stackrel{ (b, b_y)  \rightarrow 0}{\rightarrow} V_0 \delta (r ) \delta (y - y_0),
\end{equation}
whose eigenstates have only even parity~\cite{Guo:2023ecc}.

We assume that only two lowest bound states in $y$-direction are contributing, although this number can be 
manipulated in Monte Carlo simulation to be discussed next.
The two bound states are the solutions of,
\begin{equation}
\hat{H}_y^{(trap)}  \chi_n (y) = \sigma_n   \chi_n (y), \  \ n=1,2.
\end{equation}
The total 2D wavefunction can be written in the form of,
\begin{equation}
\Psi_E (r, y) = \sum_{n=1}^2 \psi_E^{(n)} (r)  \chi_n (y) ,
\end{equation}
where $n$ labels the $n$-th channel in $r$-direction.
After integrating out  dynamics along $y$-direction,  we find,
\begin{align}
&\sum_{n'=1}^2    \left [ \delta_{n,n'}  (\sigma_{n}  +  \hat{H}_r^{(\omega)}  ) + V_{n, n'} (r)  \right ] \psi_E^{(n')} (r)  = E   \psi_E^{(n)} (r)   ,
\end{align} 
where
\begin{align}
&  V_{n, n'}  (r) = \int_{y_0 - \frac{b_y}{2} }^{y_0 + \frac{b_y}{2} }  d y  \chi^*_n (y)  V(r,y)  \chi_{n'} (y)  .
\end{align}
 The quasi-1D coupled-channel dynamics can indeed be reduced from 2D systems. Hence the coupled-channel dynamics may be simulated and tested in 2D quantum mechanics by Monte Carlo methods.

\subsection{Path integral representation of quasi-1D quantum mechanical model} 
In Euclidean space-time, the path integral representation of transition amplitude for a single particle propagating from $(r,y,0)$ to $(r',y',\tau)$ is defined by,
\begin{align}
& \langle r', y' | e^{- \hat{H} \tau} | r, y \rangle = \int \prod_{i=1}^{N_{\tau}-1} d r_i d y_i e^{ - S_E^{(\omega)} (\{ r_i \}) }  \nonumber \\
& \times  \rho(y, y_1; a_{\tau}) \cdots   \rho(y_{N_\tau -1}, y' ; a_{\tau}) e^{ - S_E^{(V)} (\{ r_i, y_i \}) },
\end{align}
where the time interval $[0, \tau]$ is discretized by a one-dimensional lattice of  $N_\tau $ small steps of spacing $a_\tau = \frac{\tau}{N_\tau }$. The Euclidean space-time harmonic oscillator trapped action in $r$-direction is given by,
\begin{equation}
S_E^{(\omega)} (\{ r_i \}) = a_\tau \sum_{i=1}^{N_\tau } \left [ \frac{m}{2} \left ( \frac{ r_{i+1} -r_i }{a_\tau} \right )^2 + \frac{1}{2} m \omega^2 r_i^2 \right ],
\end{equation}
and the interaction action by,
\begin{equation}
S_E^{(V)} (\{ r_i, y_i \}) = a_\tau \sum_{i=1}^{N_\tau }  V(r_i, y_i) .
\end{equation}
The density matrix in $y$-direction is defined by,
\begin{equation}
 \rho(y_1 , y_2; a_{\tau})  = \sum_{n=1}^2 e^{- \sigma_n a_\tau} \chi_n (y_1) \chi^*_n (y_2),
\end{equation}
which obeys the relation on the lattice,
\begin{equation}
 \int \prod_{i=1}^{N_{\tau} -1}  d y_i    \rho(y, y_1; a_{\tau}) \cdots   \rho(y_{N_\tau-1}, y' ; a_{\tau})  =  \rho(y , y'; \tau) . 
\end{equation}
The spectral representation of transition amplitude is given by,
\begin{equation}
 \langle r', y' | e^{- \hat{H} \tau} | r, y \rangle   =  \sum_l   e^{- E_l \tau}  \Psi_l (r, y)  \Psi^{*}_l (r'  , y' )  ,
\end{equation}
where $E_l$ and $\Psi_l (r, y)$ are eigen-energy and eigen-wavefunction of trapped system: $\hat{H}\Psi_l (r, y) = E_l \Psi_l (r, y)$. 
In this representation, the integrated particle propagator takes the form,
\begin{equation}
 C(\tau) = \int d r d y  \langle r, y | e^{- \hat{H} \tau} | r, y \rangle = Tr   [ e^{- \hat{H} \tau}    ]   =  \sum_l   e^{- E_l \tau}    . \label{CtsumEn}
\end{equation}

The ratio of integrated particle propagators can be expressed as,
\begin{equation}
  \frac{ C(\tau)  }{C_0 (\tau)} = \int \prod_{i=1}^{N_{\tau}} d r_i d y_i  p( \{ r_i, y_i \})  e^{ - S_E^{(V)} (\{ r_i, y_i \}) }, \label{CC02Dpathinteg}
\end{equation}
where  $r_{N_\tau}   = r = r'$ and  $y_{N_\tau}   = y = y'$, and the probability density distribution is defined by,
\begin{align}
& p( \{ r_i, y_i \}) \nonumber \\
&  = \frac{ e^{ - S_E^{(\omega)} (\{ r_i \}) }    \rho(y_{N_\tau}, y_1; a_{\tau}) \cdots   \rho(y_{N_\tau-1}, y_{N_\tau} ; a_{\tau})  }{C_0 (\tau) }.
\end{align}
In this way, the interaction action factor $e^{ - S_E^{(V)} }$ and the harmonic trap action factor  $e^{ - S_E^{(\omega)} }$  are separated.
The $C_0 (\tau)$ represents the non-interacting integrated particle propagator, 
\begin{align}
  C_0 (\tau) &  = C^{(\omega)}_0 (\tau)  C^{(y)}_0 (\tau)  = Tr   [ e^{- \hat{H}_0 \tau}   ] , \label{C0tsumEn}
\end{align}
where
\begin{equation}
    C^{(\omega)}_0 (\tau)   = \int \prod_{i=1}^{N_{\tau}} d r_i   e^{ - S_E^{(\omega)} (\{ r_i \}) }    =    \sum_{l=0}^\infty e^{- \omega (l + \frac{1}{2})  \tau }  , \label{Cw0tsumEn} \end{equation}
and
\begin{align}
  C^{(y)}_0 (\tau)  & =   \int \prod_{i=1}^{N_{\tau}}  d y_i   \rho(y_{N_\tau}, y_1; a_{\tau}) \cdots   \rho(y_{N_\tau-1}, y_{N_\tau} ; a_{\tau})   \nonumber \\
& = e^{- \sigma_1 \tau } + e^{-\sigma_2 \tau}   . \label{Cy0tsumEn}
\end{align}
As a consequence, the probability density distribution is positive-definite and normalized,
\begin{equation}
  \int \prod_{i=1}^{N_{\tau}} d r_i d y_i  p( \{ r_i, y_i \})   = 1, 
\end{equation}
This property suggests that Eq.\eqref{CC02Dpathinteg} can in principle be evaluated via Monte Carlo methods.

Furthermore,  Eq.(\ref{CC02Dpathinteg}) can be simplified by integrating out dynamical degrees of freedom in  $y$-direction by defining effective potentials in $r$-direction,
\begin{equation}
\int d y_1 | \chi_n (y_1) |^2 e^{- a_\tau V (r_1, y_1) } =  e^{- a_\tau V_n (r_1)}, \ \ n =1,2, \label{V12effpotdef}
\end{equation}
and,
\begin{equation}
\int d y_1  \chi_1 (y_1)  e^{- a_\tau V (r_1, y_1) }  \chi^*_2(y_1) =  - a_\tau \gamma (r_1).  \label{Vgeffpotdef}
\end{equation}
Since wavefunctions $\chi_n (y)$ are real functions,  $V_{1,2} (r_1)$ and $\gamma (r_1)$ are real effective potentials. Due to  orthogonality of $\chi_l (y)$ functions,  the effective potentials still have square-well form,
\begin{align}
 V_n (r_1)  
&= 
\begin{cases}  
- \frac{ \ln \left [1+  \kappa_n  (e^{- \frac{a_\tau  V_0  }{b b_y}  }-1)  \right]}{a_\tau } , & if \ r_1 \in [-\frac{b}{2}, \frac{b}{2}], \\
0 , & \text{otherwise},
\end{cases} \nonumber \\
\gamma (r_1)  
&= 
\begin{cases}  
- \frac{ \kappa_\gamma (e^{- \frac{a_\tau  V_0  }{b b_y}  }-1) }{a_\tau }     , & if \ r_1 \in [-\frac{b}{2}, \frac{b}{2}], \\
0 , & \text{otherwise},
\end{cases} 
\end{align}
where the coefficients are defined by,
\begin{align}
 \kappa_n  & = \int_{ y_0 - \frac{b_y}{2} }^{  y_0 + \frac{b_y}{2}} d y_1 | \chi_n (y_1) |^2, \ \ n=1,2, \nonumber \\
  \kappa_\gamma  & = \int_{ y_0 - \frac{b_y}{2} }^{  y_0 + \frac{b_y}{2}} d y_1  \chi_1 (y_1) \chi^*_2 (y_1) .
\end{align}
The introduction  of effective potentials leads to $y$-integration that can be expressed in matrix form,
\begin{align}
& \int d y_1   \rho(y_0, y_1; a_{\tau})  e^{- a V(r_1, y_1)}  \rho(y_1, y_2; a_{\tau})  \nonumber \\
& = \begin{bmatrix} \chi_1 (y_2) & \chi_2 (y_2) \end{bmatrix} M (r_1)  \begin{bmatrix} \chi_1 (y_0) e^{- \sigma_1 a_\tau } \\ \chi_2 (y_0)  e^{- \sigma_2 a_\tau }  \end{bmatrix} ,
\end{align}
where $M (r_1)$ is defined by,
\begin{equation}
M (r_1) 
=
  \begin{bmatrix}  
  e^{- a V_1(r_1)}  e^{- \sigma_1 a_\tau }  & - a \gamma (r_1)  e^{- \sigma_1 a_\tau }   \\
- a \gamma (r_1)  e^{- \sigma_2 a_\tau } & e^{- a V_2(r_1)}  e^{- \sigma_2 a_\tau }
 \end{bmatrix}  .
\end{equation}
Carrying out all $y$-direction integrations step by step, we   find,
\begin{align}
& \int \prod_{i=1}^{N_{\tau}}  d y_i   \rho(y_{N_\tau}, y_1; a_{\tau}) \cdots   \rho(y_{N_\tau}, y_{N_\tau} ; a_{\tau}) e^{ - S_E^{(V)} (\{ r_i, y_i \}) }  \nonumber \\
&  = Tr \left [ M (r_{N_\tau })   \cdots  M (r_1)   \right ].
\end{align}
Using this property, the  path integral representation of integrated particle propagator is simplified to,
\begin{equation}
  \frac{ C(\tau)  }{C_0 (\tau)} = \int \prod_{i=1}^{N_{\tau}} d r_i  \frac{ e^{ -   S_E^{(\omega)} (\{ r_i \})  }Tr \left [ M (r_{N_\tau })   \cdots  M (r_1)   \right ] }{C_0 (\tau)} . \label{CC01Dpathinteg}
\end{equation}
This expression is reminiscent of path integrals in lattice QCD after fermion degrees of freedom are integrated out, with $Tr [ \cdots] $ serving as `fermion loops'. 
It can be evaluated by Monte Carlo importance sampling,
\begin{equation}
  \frac{ C(\tau)  }{C_0 (\tau)} = \frac{1}{N_{cfg}} \sum_{\alpha =1}^{N_{cfg}}  \frac{  Tr \left [ M (r^{(\alpha)}_{N_\tau })   \cdots  M (r^{(\alpha)}_1)   \right ] }{ e^{- \sigma_1 \tau } + e^{-\sigma_2 \tau} } , \label{CC01DMC}
\end{equation}
where $N_{cfg}$ is the total number of configurations. The random values of $\{ r_i^{(\alpha)} \}$ for the $\alpha$-th configuration are generated according to the harmonic oscillator trap probability density distribution:
 \begin{equation}
 p^{(\omega)}(\{ r_i \}) =  \frac{ e^{ - S_E^{(\omega)} (\{ r_i \}) }      }{  \int \prod_{i=1}^{N_{\tau}} d r_i  e^{ - S_E^{(\omega)} (\{ r_i \}) } } .
 \label{pomeg}
 \end{equation}

If we define the  coupled-channel propagation amplitudes matrix in Minkowski time by,
\fontsize{9}{9}
\begin{equation}
U(r, r'; t)  \stackrel{t = - i \tau}{=} \int \prod_{i=1}^{N_{\tau}-1} d r_i  e^{-  S_E^{(\omega)} (\{ r_i \})  }  M (r_{N_\tau -1})   \cdots  M (r_1)    ,
\end{equation}
\normalsize
then in the continuum limit of $a_\tau \rightarrow 0$,  the $U$ matrix  satisfies coupled-channel Schr\"odinger equation,
\begin{equation}
 i \frac{\partial U(r, r'; t) }{\partial t}  =  \left (   \begin{bmatrix} H_r^{(\omega)} & 0\\  0 & H_r^{(\omega)} \end{bmatrix}  + V(r)    \right ) U(r, r'; t) , 
\end{equation}
where $r \neq r'$, and the potential matrix approaches,
\begin{equation}
V(r) = - \frac{1}{a_\tau} \ln M(r) \stackrel{a_\tau  \rightarrow 0}{\rightarrow } \begin{bmatrix} \sigma_1+ V_1 (r) & \gamma (r) \\ \gamma (r) & \sigma_2 + V_2 (r) \end{bmatrix} .
\end{equation}
The total Hamiltonian matrix at continuum limit then takes the form,
\begin{equation}
\hat{H}^{(trap)}=    \begin{bmatrix}  \sigma_1  + H_r^{(\omega)}+ V_1 (r) & \gamma (r)  \\  \gamma (r)  & \sigma_2 +H_r^{(\omega)} + V_2 (r) \end{bmatrix}      ,
\label{Hq1d}
\end{equation}
which   indeed describes a coupled-channel system in a harmonic oscillator trap.

\subsection{Monte Carlo simulation of coupled-channel system  vs. exact solution vs. infinite volume limit}\label{MCresult} 

In this subsection, we carry out  Monte Carlo sampling calculation of $\frac{C(\tau)}{C_0 (\tau)}$ defined in Eq.(\ref{CC01DMC}) with two different traps in $y$-direction: a harmonic trap or a hard wall trap. In both cases, two lowest even-parity states are installed in the density matrix. Since the eigenstates are analytical in both cases, they lead to exact effective potentials in $r$-direction, which offers a way to check the Monte Carlo. For harmonic oscillator trap with angular frequency $\omega_y$, 
\begin{align}
\chi_1 (y)  & =  \left ( \frac{ m \omega_y}{\pi} \right )^{\frac{1}{4}} e^{- \frac{ m \omega_y}{2} y^2} H_0 (\sqrt{ m \omega_y} y)  , \ \  \sigma_1 = \frac{\omega_y}{2} , \nonumber \\
\chi_2 (y)  & = \frac{1}{\sqrt{2^3}} \left ( \frac{ m \omega_y}{\pi} \right )^{\frac{1}{4}} e^{- \frac{ m\omega_y}{2} y^2} H_2 (\sqrt{ m \omega_y} y), \ \  \sigma_2 = \frac{5 \omega_y}{2}. \label{HOtrapstateiny}
\end{align}
For hard-wall trap with walls placed at the two boundaries $y=0$ and $y=L_y$, 
\begin{align}
\chi_1 (y)  & =   \sqrt{\frac{2}{L_y}} \sin \frac{\pi y}{L_y}  , \ \  \ \  \sigma_1 = \frac{\pi^2}{2m L_y^2} , \nonumber \\
\chi_2 (y)  & =  \sqrt{\frac{2}{L_y}} \sin \frac{3 \pi y}{L_y}  , \ \  \ \  \sigma_2 = \frac{9 \pi^2}{2m L_y^2} . \label{HWtrapstateiny}
\end{align}

The quasi-1D coupled-channel effective potentials have square-well shape:
\begin{align}
 V_n (r)  
&= 
\begin{cases}  
 \frac{V_n}{b}   , & if \ r \in [-\frac{b}{2}, \frac{b}{2}], \\
0 , & \text{otherwise},
\end{cases}, \ \ n=1,2,   \nonumber \\
 \gamma  (r)  
&= 
\begin{cases}  
   \frac{\gamma}{b}     , & if \ r \in [-\frac{b}{2}, \frac{b}{2}], \\
0 , & \text{otherwise}.
\end{cases}  \label{VGeffpotquasi1D}
\end{align}
The strengths of the square well potentials are given by
\begin{align}
 V_n    
&= 
 - \frac{b}{a_\tau } \ln \left [1+  \kappa_n  (e^{- \frac{a_\tau  V_0  }{b b_y}  }-1)  \right]  , \ \ n=1,2,  \nonumber \\
 \gamma  
&= 
  - \frac{b}{ a _\tau } \kappa_\gamma (e^{- \frac{a_\tau  V_0  }{b b_y}  }-1)      ,    \label{strengthVGeffpotquasi1D}
\end{align}
where the $\kappa$ parameters  are, for harmonic oscillator trap ($y_0 = 0$),
\begin{align}
 \kappa_1  & =   \mbox{erf} ( \frac{\sqrt{m \omega_y b_y^2} }{2}) , \nonumber \\
 \kappa_2 & =  \mbox{erf}  ( \frac{\sqrt{m \omega_y b_y^2} }{2}) -   \frac{ 2 + m\omega_y b_y^2}{4}  \sqrt{ \frac{m \omega_y b_y^2}{\pi} }  e^{- \frac{m \omega_y b_y^2 }{4} }  , \nonumber \\
  \kappa_\gamma  & =  -  \sqrt{ \frac{m \omega_y  b_y^2}{2\pi} } e^{- \frac{m \omega_y b_y^2}{4} } ;
\end{align}
and for hard wall trap ($y_0 = \frac{L_y}{2}$),  
\begin{align}
 \kappa_1  & =    \frac{b_y}{L_y}  +  \frac{1   }{ \pi} \sin \frac{ \pi b_y }{L_y}   , \nonumber \\
 \kappa_2 & =    \frac{b_y  }{L_y}  +  \frac{1   }{3 \pi} \sin \frac{3 \pi b_y }{L_y}  , \nonumber \\
  \kappa_\gamma  & =  -  \frac{  1  + \cos \frac{\pi b_y }{L_y}  }{\pi}  \sin \frac{ \pi b_y}{L_y}   .
\end{align}

 With all the ingredients in place, we are ready to make a detailed comparison 
in the quasi-one-dimensional model for coupled-channel dynamics from three perspectives.

First, the Monte Carlo simulation of  $\frac{C (\tau) }{ C_0 (\tau) }   $ in Eq.(\ref{CC01DMC}) for quasi-1D coupled-channel systems in a harmonic oscillator trap along $r$-direction can be  carried out by either standard Metropolis algorithm  \cite{Creutz:1980gp,Lepage:1998dt} or Hybrid Monte Carlo (HMC) algorithm \cite{DUANE1987216} (HMC is used in this work).  The Monte Carlo sampling configurations for non-interacting particles in a harmonic oscillator trap  are generated according to Eq.\eqref{pomeg}.
The simulation is performed with fixed lattice spacing $a_\tau \sim 0.04$, so the   number of steps in temporal dimension, $N_\tau$  varies for   $\tau \in  [0.5, 5.5]$.  Typically a half million measurements are generated for each $\tau$.  The  choice of other parameters are  $V_0 =1$, $m =1$ and $b = b_y=0.2$ for a square well potential. 
Three different sizes ($\omega  = 0.1,0.2$ and $0.5$) are used  for the harmonic oscillator trap  in $r$-direction.

Second,  the energy spectrum $\{E_l\}$ of  quasi-1D coupled-channel systems  in a harmonic oscillator trap  interacting with  square well potentials defined in Eq.(\ref{VGeffpotquasi1D}) and Eq.(\ref{strengthVGeffpotquasi1D})  can be obtained by diagonalizing the Hamiltonian in Eq.(\ref{HmatHOtrap}), 
where the matrix elements are evaluated by
 \begin{align}
 V^{(n)}_{l,l'}  &= V_n \int_{-\frac{b}{2}}^{\frac{b}{2}}  d r \phi_l ( r )        \phi_{l'} (r )  ,  \ \ n=1,2,   \nonumber \\
 \gamma_{l,l'}  &= \gamma \int_{-\frac{b}{2}}^{\frac{b}{2}} d r \phi_l ( r )       \phi_{l'} (r )  ,
 \end{align}  
using the  harmonic oscillator basis in Eq.\eqref{eq:ho-basis}.
The exact solution of  $\frac{C (\tau) }{ C_0 (\tau) } $ can then be obtained by feeding $\{E_l\}$ to $C(\tau )$ in Eq.\eqref{Ctau2}, and using $C_0(\tau )$ in Eq.\eqref{Czero2}.

Third,   the scattering solutions in infinite volume can be well approximated by contact interactions for small $b$.  The analytic expression of phase shifts and inelasticity for contact interaction potentials are given in Eq.(\ref{Tscatamps}) and Eq.(\ref{phaseinelastexpression}). At infinite volume limit,  
 \begin{equation}
 \frac{C(\tau) }{C_0 (\tau)}  \stackrel{\omega \rightarrow 0}{\rightarrow} 1+   \frac{1}{C_0 (\tau)}  \frac{\tau}{\pi}  \left [ \int_{\sigma_1}^{\infty}   \delta_1(\epsilon)   + \int_{\sigma_2}^{\infty}      \delta_2(\epsilon)  \right ]  e^{- \epsilon \tau}  d \epsilon .   \label{CoverC0inflimit}
 \end{equation} 
 It is basically Eq.\eqref{mainresult} in  a harmonic trap recast in ratio form for comparison purposes.

  \begin{figure}
 \centering
 \includegraphics[width=0.95\textwidth]{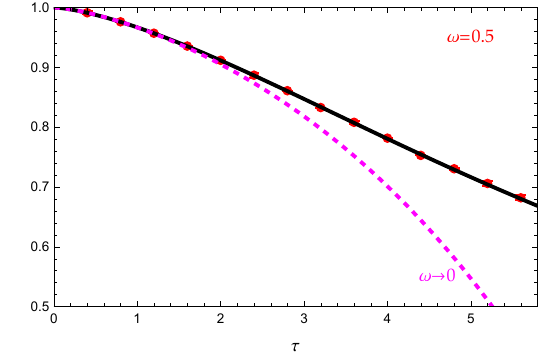}
\includegraphics[width=0.95\textwidth]{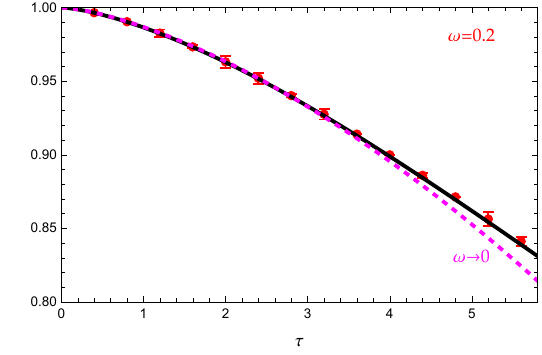}
\includegraphics[width=0.95\textwidth]{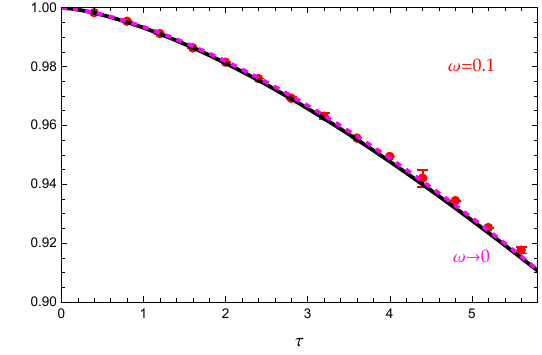}
 \caption{Convergence check of Eq.\eqref{CoverC0inflimit} for {\bf harmonic oscillator} trap placed in $y$-direction.
 The three graphs correspond to harmonic oscillator trap size $\omega =0.1,0.2,0.5$ in $r$-direction.
 The red points with error bars are  Monte Carlo data of $\frac{C(\tau)}{C_0 (\tau)}$ on the left-hand-side. 
The solid black curves are exact results for $\frac{C(\tau)}{C_0 (\tau)}$ from energy spectrum representation.   
The dashed purple curves are the infinite volume limit on the right-hand-side  of Eq.\eqref{CoverC0inflimit}. 
Other parameters are taken as:  $V_0 =5$, $b=0.2$, $a_\tau=0.04$, $m =1 $, $\sigma_1 = 0.1$, and $\sigma_2 =0.5$.  The effective potential strengths are  $V_1=0.257$, $V_2 =0.126$, and $\gamma=-0.177$.}    \label{ctoverc0tHOplots}
 \end{figure} 
  \begin{figure}
 \centering
 \includegraphics[width=0.95\textwidth]{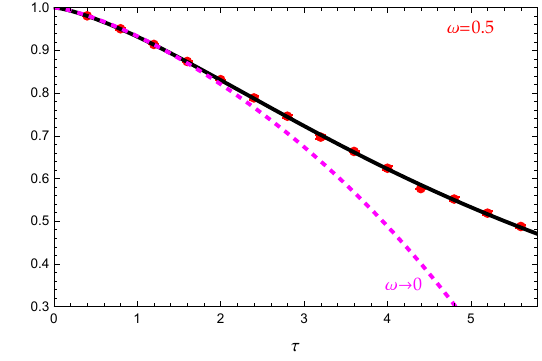}
\includegraphics[width=0.95\textwidth]{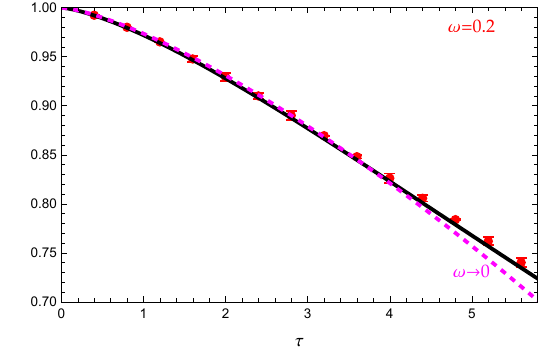}
\includegraphics[width=0.95\textwidth]{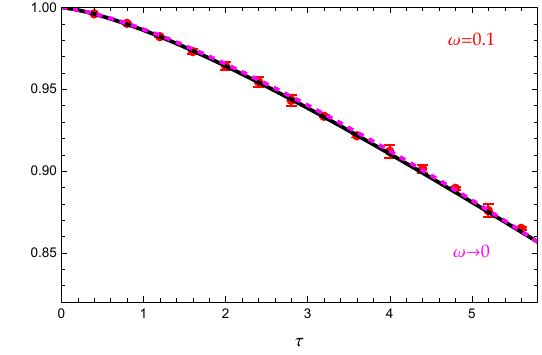}
 \caption{
 Similar to Fig.~\ref{ctoverc0tHOplots} but for {\bf hard-wall} trap placed in $y$-direction with parameters $L_y = 4$, $y_0=2.0$ and $b_y=0.2$.
Other parameters are taken as:  $V_0 =5$, $b=0.2$, $a_\tau=0.04$, $m =1 $, $\sigma_1 = 0.31$, and $\sigma_2 =2.77$.  The effective potential strengths are     $V_1=0.522$, $V_2 =0.513$, and $\gamma=-0.492$. 
 }   
 \label{ctoverc0tHWplots} 
 \end{figure}

In Fig.~\ref{ctoverc0tHOplots} we show the comparison for the three sizes of harmonic oscillator trap in $r$-direction  and harmonic oscillator trap in $y$-direction with parameters of $\omega_y=0.2$ and $y_0=0$.
The same is shown in Fig.~\ref{ctoverc0tHWplots} for hard wall trap in $y$-direction with parameters of $L_y=4$ and $y_0= 2$.
We see that the  Monte Carlo results are in excellent agreement with the exact solutions, and they both approach their infinite volume limit rapidly as the size of trap in $r$-direction is widened ($\omega \rightarrow 0$).
Moreover, the agreement is insensitive to the type of traps used in $y$-direction.
Validated by exact solutions, the Monte Carlo approach can serve as a general method to the coupled-channel formalism where  general interaction and trap potentials can be used.

\vspace{2cm}
\section{Summary and Outlook}\label{summary}

In this work, we extended the integrated correlation function formalism developed in Refs.~\cite{Guo:2023ecc,Guo:2024zal} from single channel to two coupled channels.
The central result in Euclidean space is represented by Eq.\eqref{mainresult}. Remarkably, it bears the same structure as the single-channel result in Eq.\eqref{single2}: just from a single phase shift term to sum of two phase shift terms with appropriate thresholds on the right hand side.
Both relations are reproduced here for a side-by-side comparison.

\vspace{2mm}
{\bf Single channel:} \\
$$  C(\tau) - C_0 (\tau) \stackrel{\text{trap} \rightarrow \infty}{\rightarrow} \frac{\tau}{\pi} \int_0^{\infty} d \epsilon  \,\delta(\epsilon)\, e^{- \epsilon \tau}.$$

{\bf Two coupled channels:} \\
$$ C(\tau) - C_0 (\tau) \stackrel{\text{trap} \rightarrow \infty}{\rightarrow} \frac{\tau}{\pi} \left [ \int_{\sigma_1}^{\infty}   \delta_1(\epsilon)   + \int_{\sigma_2}^{\infty}     \delta_2(\epsilon)  \right ]  e^{- \epsilon \tau}  d \epsilon. $$

In this formalism, the difference of integrated correlation functions between interacting and non-interacting systems in a trap is related to the infinite-volume phase shifts by a weighted integral over energy.  
It can be regarded as an alternative method to the well-known L\"{u}scher formula.
The new relation is validated in a number of ways, including exactly solvable models and Monte Carlo simulation that can admit general interaction potentials.
Several comments are in order.

\begin{enumerate}[leftmargin=*]
\item
The relation has two salient features. One is the rapid convergence to the infinite-volume limit at short times  as the trap size is increased. The other is that it involves directly correlation functions, bypassing the requirement to extract energy spectrum in traditional methods. 
Both features bode well for application in lattice QCD simulations to overcome signal-to-noise and energy level density issues often encountered in baryon systems as the volume is increased.

\item
Given the integrated correlation functions, we only have one constraint for two phase shifts via the relation. The situation is similar to the coupled-channel L\"uscher and   BERW formalism, see e.g. Refs.\cite{Guo:2012hv,Guo:2013vsa,Guo:2021uig}, where the quantization condition involves two phase shifts and one inelasticity. One strategy of extracting coupled-channel dynamics is to model the system based on chiral perturbation theory or dispersion relation with a few free parameters, so that the explicit dependence on these parameters and expressions for the phase shifts and inelasticity are available. By fitting the two phase shifts to the integrated correlated function data, the free parameters of the model may be determined, which in turn yields the inelasticity. Alternatively, extra constraints may be obtained by defining the correlation functions in each individual channel. This discussion is detailed in Appendix~\ref{Sum:inelasticeffect}.

\item 
Although developed for two coupled channels, the relation straightforwardly carries over to more than two channels,
\begin{equation} 
C(\tau) - C_0 (\tau) \stackrel{\text{trap} \rightarrow \infty}{\rightarrow} \frac{\tau}{\pi} \sum_{n >2} \int_{\sigma_n}^{\infty}   \delta_n(\epsilon) e^{-\epsilon \tau} d \epsilon. 
\end{equation} 
The reason is that the coupled-channel relation is only explicitly related to phase shifts, not inelasticity ($\eta$). The key observation is that the integrated correlation function of coupled-channel systems in infinite volume is defined through the trace of full Green's function matrix, see Eq.(\ref{integratedCdef}). 
For multi-channel systems, the trace of Green's function is related to the determinant of $S$-matrix~\cite{Guo:2024bar} by, 
\fontsize{9}{9} 
\begin{equation} 
 Tr  \left [ \mathcal{G}^{(\infty)} (  E) -\mathcal{G}_0^{(\infty)} (    E)   \right ]    = -  \frac{1}{\pi}  \int_{0}^{\infty} d \lambda  \frac{ \frac{1}{2 i}\ln \det \left [S(\lambda) \right] }{ (\lambda - E)^2}, 
\end{equation}
\normalsize
where the determinant is only explicitly dependent on the phase shifts,
\begin{equation}
\det \left [S(\lambda) \right]  = e^{2 i \sum_n \delta_n (\lambda)}.
\end{equation}
The inelasticities are implicitly present in the formalism.

\item
We envision no issues in extending the formalism to relativistic dynamics as demonstrated in the single-channel case~\cite{Guo:2024zal}.

\item
The trap in the formalism refers to any potential that can confine the system to have quantized energy levels. 
The common choices are harmonic oscillator trap, hard-wall trap, periodic box of size $L$, or periodic lattices of size $N$ and spacing $a$ that extrapolate to $\lim_{\stackrel{N\to\infty}{a\to 0}} N a = L$.
Future work would benefit from applying the formalism in more practical scenarios.
One possible avenue is to build on the $\phi^4$ lattice model in Ref.\cite{Guo:2024zal} to go from 1+1 dimensions  to 3+1 dimensions. Such a demonstration can inform us on the convergence rate and computational demand in Monte Carlo simulations of lattice field theories.

\end{enumerate}

\acknowledgments
This research is supported in part  by the U.S. National Science Foundation under Grant No. NSF PHY-2418937 (P.G.) and NSF PHY-1748958 (P.G.), and the U.S. Department of Energy under Grant No. DE-FG02-95ER40907 (F.L.).

\bibliography{ALL-REF.bib}

\appendix

\section{Inelastic effect in individual channel of a coupled-channel system}\label{Sum:inelasticeffect}

The inelastic effect  in total     integrated correlation functions by summing over the contribution from both channels is not explicitly present in Eq.(\ref{mainresult}) via inelasticity. As discussed in Ref.~\cite{Guo:2024bar}, the imaginary part of the trace of integrated Green's function is related to the determinant of $S$-matrix by
\begin{equation}
Im  \left [ Tr \left [ G^{(\infty) } (\lambda) -  G_0^{(\infty) } (\lambda) \right ] \right ] = - \frac{1}{2 i} \frac{d}{d \lambda} \ln  \det \left [S(\lambda) \right ],
\end{equation}
where 
\begin{equation}
    \det \left [S(\lambda) \right ] = e^{2 i [ \delta_1 (\lambda) + \delta_2 (\lambda) ] }.
\end{equation}
 The inelasticity is canceled out in determinant of $S$-matrix, which is warranted by  the  Friedel formula, see e.g. Refs.~\cite{Faulkner_1977,Guo:2022row}.  We show in the following that the explicit dependence of inelasticity  shows up in each individual channel integrated correlation functions.

As illustrated in Ref.~\cite{Guo:2024bar}, for each individual channel,  the  diagonal terms of infinite volume integrated Green's function for coupled-channel systems are 
\begin{align}
&  \int_{-\infty}^\infty d r \left [ G^{(\infty)}_{nn} (r,r; E) - G_n^{(\infty,0)} (0;E) \right ] \nonumber \\
&  =  - \frac{\mu_n}{k_n} \frac{\partial}{\partial k_n} \ln t_{nn} (k_1, k_2), \label{integratedGii}
\end{align}
where $t_{nn} (k_1, k_2)$ denote diagonal terms of transmission amplitude that are defined via diagonal  scattering amplitudes by
\begin{equation}
t_{nn} (k_1, k_2) = 1+ i \frac{\mu_n}{k_n} T_{nn}  = \frac{\eta e^{2 i \delta_n } +1}{2} .
\end{equation}
Notation-wise, the dependent on $(k_1,k_2)$ in $t_{nn} (k_1, k_2)$ is written explicitly to remind readers that the diagonal term of integrated Green's function is related to corresponding transmission amplitude by partial derivative. We also find
 \fontsize{9}{9} \begin{equation}
Im  \int_{-\infty}^\infty d r \left [ G^{(\infty)}_{nn} (r,r; E) - G_n^{(\infty,0)} (0;E) \right ]  =  - \frac{\mu_n}{k_n} \frac{\partial  \Phi_n (E)}{\partial k_n} , \label{ImGii}
\end{equation} \normalsize
 where $ \Phi_n (E)$ is the phase of $t_{nn} (k_1, k_2)$:
 \fontsize{9}{9} \begin{equation}
 \Phi_n (E)  = \tan^{-1} \frac{Im [ t_{nn} (k_1, k_2) ]}{Re [ t_{nn} (k_1, k_2) ]} =  \tan^{-1} \left [ \frac{\eta \sin (2 \delta_n )}{\eta \cos (2\delta_ i ) +1} \right ]. \label{Phiphasedef}
 \end{equation} \normalsize
 At the elastic limit by turning off coupling $\eta \rightarrow 1$, $ \Phi_n (E)$ is reduced to the elastic scattering phase shift $ \delta_n (E)$.  Let's define a extra phase factor by
  \begin{equation}
 \Delta_n (E) = \delta_n (\sigma_n) + \int_{\sigma_n}^E d \lambda   \frac{\mu_n}{k_n (\lambda)} \frac{\partial  \Phi_n (\lambda)}{\partial k_n (\lambda)} , \label{Deltafunctiondef}
 \end{equation}
 so that
  \begin{equation}
 \frac{d  \Delta_n (E)  }{d E}  =    \frac{\mu_n}{k_n} \frac{\partial  \Phi_n (E)}{\partial k_n } ,
 \end{equation}
 hence Eq.(\ref{ImGii}) can be written as
 \fontsize{9}{9}
 \begin{equation}
Im  \int_{-\infty}^\infty d r \left [ G^{(\infty)}_{nn} (r,r; E) - G_n^{(\infty,0)} (0;E) \right ]  =  - \frac{d   \Delta_n (E) }{d E}   .   \label{ImGiidDeltadE}
\end{equation}
\normalsize
Using the analytical properties of Green's function, Eq.(\ref{integratedGii}) is thus given in terms of $\Delta_n (E) $ by
\begin{align}
&  \int_{-\infty}^\infty d r \left [ G^{(\infty)}_{nn} (r,r; E) - G_n^{(\infty, 0)} (0;E) \right ]  \nonumber \\
&  =  - \frac{1}{\pi} \int_{\sigma_n}^\infty d \lambda \frac{\Delta_n (\lambda) }{(\lambda - E)^2}  . \label{GiidDeltadE}
\end{align}
Therefore we find
\begin{align}
& \int_{-\infty}^\infty \frac{d\lambda}{2\pi}    \int_{-\infty}^\infty d r \left [ G^{(\infty)}_{nn} (r,r; E) - G_n^{(\infty,0)} (0;E) \right ]  e^{- i \lambda t}  \nonumber \\
 &=   \frac{t}{\pi} \int_{\sigma_n}^\infty d \epsilon   \Delta_n (\epsilon)  e^{- i \epsilon t}  . \label{Ginffriedelrelation}
\end{align}
Comparing Eq.(\ref{ImGinfphaseshift}) with Eq.(\ref{ImGiidDeltadE}), we obtain 
 \begin{equation}
   \sum_{i =1}^2 \frac{d   \Delta_n (E) }{d E}   = \frac{d  [  \delta_1 (E) + \delta_2 (E) ] }{d E}    .    \label{sumdDeltadE}
\end{equation}
In fact, with some tedious calculation and using relations given in Ref.~\cite{Guo:2024bar}, we can show that
\fontsize{9}{9} \begin{equation}
\sum_{n=1}^2 \frac{\mu_n}{k_n} \frac{\partial}{\partial k_n} \ln t_{nn} (k_1, k_2) =  \frac{d}{d E} \left [  \frac{1}{\pi} \int_{\sigma_1}^\infty d \lambda \frac{\delta_1 (\lambda) + \delta_2 (\lambda)}{\lambda - E}  \right ],
\end{equation} \normalsize
the imaginary part of above relation therefore yields the relation in Eq.(\ref{sumdDeltadE}).

\begin{figure*}
 \centering
\includegraphics[width=0.48\textwidth]{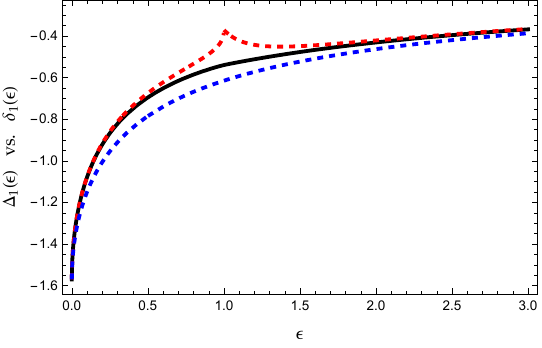}
\includegraphics[width=0.48\textwidth]{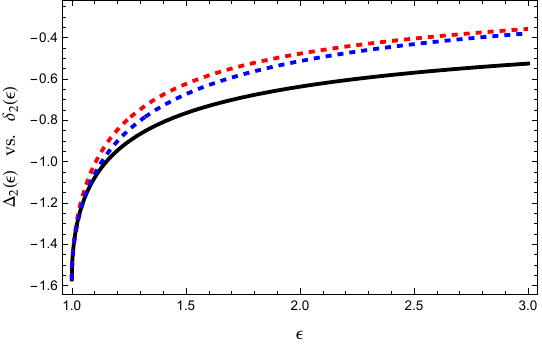}
 \caption{Comparison of phase shifts in channel 1 (left) and channel 2 (right).  The three curves correspond to: $\Delta_{n} (\epsilon)$ (solid black), $\delta_{n} (\epsilon)$ (dashed red), and $-\cot^{-1} ( \frac{k_n}{\mu_n V_n})$ (dashed blue).   The  parameters are taken as:       $V_1=1.0$, $V_2 =0.8$, $\gamma=0.6$, $\sigma_1 =0$, $\sigma_2 =1$ and $\mu_1 = \mu_2 =1$.}   \label{Deltaphaseplots} 
 \end{figure*}

For the trapped systems, if the integrated correlation function for each individual channel is defined by
\begin{equation}
C_{n} (t) = i \int_{-\infty}^\infty \frac{d \lambda}{2\pi}   \left [    \int d r G_{nn}^{(trap)} (r,r;  \lambda + i 0) \right ] e^{- i \lambda t} , \label{integratedCiidef}
\end{equation} 
using Eq.(\ref{Ginffriedelrelation}), we find
 \begin{equation}
 C_{n}(t) - C_n^{(0)} (t) \stackrel{\text{trap} \rightarrow \infty}{\rightarrow} \frac{i \, t}{\pi}  \int_{\sigma_n}^{\infty}   \Delta_n (\epsilon)     e^{- i \epsilon t}  d \epsilon . \label{dCiiresult}
 \end{equation} 
 The inelastic effect for the single channel is embedded in the $  \Delta_n (E)  $ functions, which is related to both phase shifts and inelasticity through  Eq.(\ref{Deltafunctiondef}) and Eq.(\ref{Phiphasedef}).

Using the contact interaction model discussed in Sec.~\ref{exactmodel} as a specific  example, we can also show that
\begin{align}
&  \int_{-\infty}^\infty d r \left [ G^{(\infty)}_{nn} (r,r; E) - G_n^{(\infty,0)} (0;E) \right ] \nonumber \\
&  =   i \frac{\mu^2_n}{k^3_n}  T_{nn}  =  \frac{\mu_n}{k^2_n}    \frac{\eta e^{2 i \delta_n} -1}{2},
\end{align}
where  the analytic expression of $T_{nn}$'s are given in Eq.(\ref{Tscatamps}),   therefore
\begin{equation}
     \frac{\mu_n}{k_n} \frac{\partial  \Phi_n (E)}{\partial k_n}  =  - Re \left [ \frac{\mu^2_n}{k^3_n}  T_{nn}   \right ]  =  -  \frac{\mu_n}{k^2_n}   \frac{\eta \sin (2\delta_n )}{2}, 
\end{equation}
and $ \Delta_n (E)$ can be computed by
 \begin{equation}
 \Delta_n (E) = \delta_n (\sigma_n) - \int_{\sigma_n}^E d \lambda    \frac{\mu_n}{k^2_n (\lambda)}   \frac{\eta (\lambda) \sin (2\delta_n (\lambda) )}{2}.  
 \end{equation}
As $\gamma \rightarrow 0$, the two channels are completely decoupled, leading to,
  \begin{equation}
 \Delta_n (E)  \stackrel{\gamma \rightarrow 0}{  \rightarrow }  \delta_n (E)    \rightarrow   - \cot^{-1} \left ( \frac{k_n}{ \mu_n V_n} \right )   ,  
 \end{equation}
 in each channel.
 The comparison of $ \Delta_n (E) $ vs. $ \delta_n (E) $ vs. $ - \cot^{-1} \left ( \frac{k_n}{ \mu_n V_n} \right ) $ is demonstrated  in Fig.~\ref{Deltaphaseplots}. Significant effects of inelasticity can be observed in both channels as the difference between the solid and dashed curves.

\end{document}